\documentclass[12pt,a4paper]{article}
\usepackage{epsfig}
\usepackage{graphics}
\usepackage{amsmath}

\newcommand{\bmath}[1]{\mbox{\boldmath{${#1}$}}}

\newcommand{\half}{\mbox{${\textstyle \frac{1}{2}}$}}           
\newcommand{\third}{\mbox{${\textstyle \frac{1}{3}}$}}          

\newcommand{\rd}{\textrm{d}}

\begin{document}
\setlength{\unitlength}{1mm}
%
\newpage
\hspace{15cm}

\begin{flushright}
August 2, 2007
\end{flushright}

\vspace{5mm}

\begin{center}

\vspace*{10mm}

{\Large Hard pion bremsstrahlung in the Coulomb region }
\\[10ex]
G{\"o}ran F{\"a}ldt\footnote{E-mail: goran.faldt@tsl.uu.se} 
  and Ulla Tengblad\footnote{E-mail: ulla.tengblad@tsl.uu.se}  
\\[3ex]
Department of nuclear and particle physics\\
Uppsala University\\
 Box 535, S-751 21 Uppsala \\[8ex]

\vspace{5mm}

{\bf Abstract}
\end{center}
Hard high-energy pion-nucleus bremsstrahlung, $
\pi^- +A\rightarrow\pi^- +\gamma +A$,
is studied in the Coulomb region, {\it i.e.}~the small-angle
region where the nuclear scattering is dominated by
the Coulomb interaction. Special attention is focussed on the
possibility of measuring the pion polarizability in such reactions. 
We study the sensitivity to the structure of the underlying 
pion-Compton amplitude
through a model with $\sigma$, $\rho$, and $a_1$ 
exchanges. It is found that the effective energy in the
virtual pion-Compton scattering is often so large
that the threshold approximation does not apply.   

\vspace{5mm}
\noindent

\vfill

\newpage
%
%
%
\section{Introduction}

High-energy pionic bremsstrahlung, {\it i.e.}~the coherent nuclear reaction
\[
\pi^- +A\rightarrow\pi^- +\gamma +A
\]
can proceed through a one-photon exchange. In fact, at small momentum transfers 
to the nucleus $A$, the reaction is dominated by the virtual pion-Compton reaction
$\gamma+\pi^- \rightarrow \gamma+\pi^-$.
A long time ago it was suggested \cite{3,4}, that by studying pionic bremsstrahlung
 important information on the pion-Compton amplitude could be extracted. Of 
 particular interest is the pion electric and magnetic polarizabilities, 
 which are low-energy parameters that have been calculated in chiral-Lagrangian
 theory \cite{7}.
 A bremsstrahlung experiment  aiming at measuring the polarizabilities  
 has been performed  \cite{5}, and  reasonable values 
 of these parameters were extracted. The pion polarizabilities can also be
 determined in other reactions, such as pion photoproduction \cite{5a}.
 
At low energies the pion-Compton amplitude
can be regarded as a sum of two contributions,  a structure-independent 
Born term, and a  structure-dependent term fixed by the pion
polarizabilities. At higher energies the situation is more complex.
Therefore, we have chosen to model the pion-Compton amplitude 
as a sum of the Born amplitudes, and the amplitudes generated by the
$\sigma$, $\rho$, and $a_1$ exchanges. This model should be fairly reliable 
also in the early  GeV region.

In a previous paper \cite{PrimakI} we developed a Glauber model for pion-nucleus 
bremsstrahlung. Such a model includes nuclear scattering 
and is also valid for momentum transfers outside the Coulomb region of small momentum
transfers. For the pion-Compton amplitude only the Born terms
and the polarizabilities were retained. However, it was pointed
out that in applications one quickly comes into a region of
high energies in the virtual pion-Compton scattering. 
In the present paper we direct our interest at exactly this 
energy dependence. The meson-exchange model is then the reasonable 
starting point. Furthemore, we consider only the small-angle region
where it is sufficient to retain the Coulomb interaction alone, and 
neglect all nuclear interactions.
%
%
%
\newpage
\section{Pion-Compton scattering} 

The primary mechanism responsible for pion-nucleus bremsstrahlung in the Coulomb
region is pion-Compton scattering, involving a virtual  photon exchange 
between the pion and the nucleus. 
In our previous investigation 
\cite{PrimakI} we used the low-energy approximation of the 
pion-Compton amplitude,
as parametrised by the pion polarizabilities. Now, we want to go 
beyond this approximation, and investigate, in a model, the limits
of the low-energy approximation in actual applications. 
We shall assume that, in addition to the Born terms,
the pion-Compton amplitude receives contributions also from
 the $\sigma$, $\rho$ and  $a_1$ exchange diagrams.

The Compton amplitude is written as
\[
{\mathcal M}(\gamma(q_1)\pi^-(p_1)\rightarrow \gamma(q_2)\pi^-(p_2))=
  {\mathcal M}_{\mu\nu}\epsilon_1^{\mu}(q_1)\epsilon_2^{\nu}(q_2)\ .
\]
Gauge invariance requires that, for real as well as virtual photons
with  $q^2\neq0$, the Compton tensor satisfies 
\[
  {\mathcal M}_{\mu\nu}q_1^{\mu}= {\mathcal M}_{\mu\nu}q_2^{\nu}=0 \ .
\]
The Compton tensor ${\mathcal M}_{\mu\nu}$ is conveniently 
decomposed as
\begin{equation}
	{\mathcal M}_{\mu\nu}=ie^2 \left[ A(s,t) {\mathcal A}_{\mu \nu}
	   + B(s,t) {\mathcal B}_{\mu \nu} \right]   \ ,   \label{Compt-decomp}
\end{equation}
with the gauge-invariant tensors ${\mathcal A}_{\mu \nu}$ and 
${\mathcal B}_{\mu \nu}$ defined as
\begin{eqnarray}
  {\mathcal A}_{\mu\nu}&=& 2g_{\mu\nu}
  -\frac{(2p_2+q_2)_{\nu}(2p_1+q_1)_{\mu}}{s-m_{\pi}^2}
  -\frac{(2p_1-q_2)_{\nu}(2p_2-q_1)_{\mu}}{u-m_{\pi}^2} \ ,
\label{A-tensor}\\ 
 {\mathcal B}_{\mu\nu}&= & q_1\cdot q_2\, g_{\mu\nu} -q_{2\mu} q_{1\nu}   \ ,
\label{B-tensor}
\end{eqnarray}  
and the Mandelstam kinematic variables by
\begin{equation}
\begin{array}{rcl}
  s&=& (p_2+q_2)^2 \ ,\\
  t &=& (p_1-p_2)^2 \ ,\\ 
  u&=& (p_1- q_2)^2 \ .
\end{array}    
\label{Mandelstam}
\end{equation}  

For pions there are three Born amplitudes described by the 
Feynman diagrams of fig.~1. In the decomposition of 
eq.(\ref{Compt-decomp})
the invariant functions $ A(s,t)$ and $ B(s,t)$ are 
\begin{eqnarray}
  A(s,t)&=& 1 \ , \label{A-facB}\\ 
 B(s,t)&= & 0  \ .  \label{B-facB}
\end{eqnarray} 
\begin{figure}[t]
\begin{tabular}{c@{}c@{}c@{}}
\scalebox{1.0}{\includegraphics{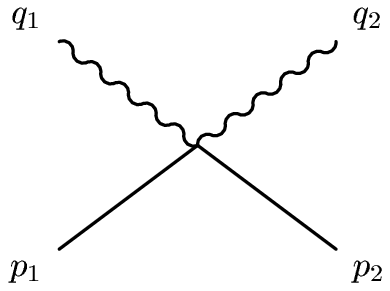}}&\quad
\scalebox{1.0}{\includegraphics{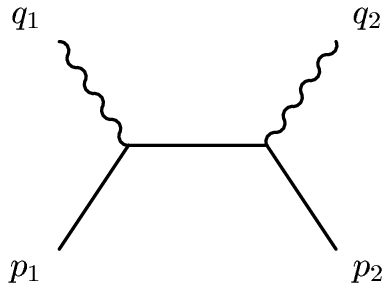}}&\quad
\scalebox{1.0}{\includegraphics{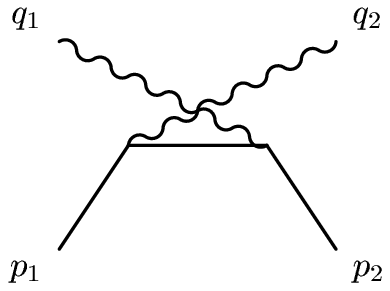}}  
 \end{tabular}
 \caption{Born diagrams for pion-Compton scattering.}
  \label{fig:1}
\end{figure}\\

In our previous study we went beyond the Born approximation,
adding the threshold contributions represented by  the electric 
and magnetic polarizabilities, $\alpha_{\pi}$ and $\beta_{\pi}$, 
leading to the result 
\begin{eqnarray}
  A(s,t)&=& 1 +\frac{\alpha_{\pi} +\beta_{\pi}}{4 m_{\pi} \alpha} 
       (s-m_{\pi}^2) (u- m_{\pi}^2)  \label{A-fac0} \ , \\ 
 B(s,t)&= & \frac{2 m_{\pi}\beta_{\pi}}{\alpha}  \ . \label{B-fac0}
\end{eqnarray}  
In chiral-Lagrangian theory \cite{7} numerical values 
are $\alpha_{\pi}+\beta_{\pi}=0$
and $\alpha_{\pi}=2.7\cdot10^{-4}$ fm$^3$.

A model for the energy dependence of the Compton amplitude
can be obtained by invoking, in addition to the Born terms,
 the contributions from the $\sigma(0^+)$, $\rho(1^-)$ and 
$a_1(1^+)$ exchanges  graphed in fig.~2.
\begin{figure}[ht]
\begin{tabular}{c@{}c@{}c@{}}
\scalebox{1.10}{\includegraphics{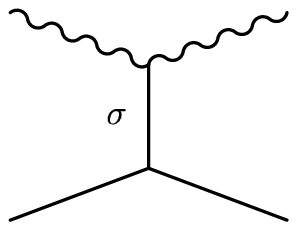}}&\quad
\scalebox{1.10}{\includegraphics{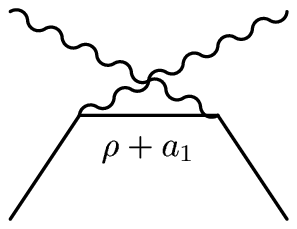}} &\quad
 \scalebox{1.10}{\includegraphics{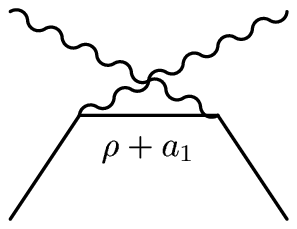} }
 \end{tabular}
  \caption{Feynman diagrams for the $\sigma$, $\rho$ and 
$a_1$ contributions to pion-Compton scattering.}
\end{figure}\\
Such models have been investigation in connection with 
studies of the reaction $\gamma\gamma\rightarrow\pi\pi$,
and its $t$-channel counterpart 
$\gamma\pi\rightarrow\gamma\pi$. Numerical values of the 
model parameters have been extracted from experimental data
by Fil'kov and Kashevarov \cite{Fil}.

The evaluation of the diagrams of fig.~2 is straightforward. 
We parametrise the Compton amplitude through dimensionless 
functions $\lambda_1(s,t)$ and $\lambda_2(s,t)$ 
rather than $\alpha_{\pi} +\beta_{\pi}$ and $\beta_{\pi}$. 
Thus we introduce for the invariant functions 
$A(s,t)$ and $ B(s,t)$ of eq.(\ref{Compt-decomp}) the decompositions
\begin{eqnarray}
	A(s,t) &=& 1+\frac{(s-m_{\pi}^2)(u-m_{\pi}^2)}{4m_{\pi}^4} \lambda_1(s,t)
     \ , \label{A-coeff} \\
  B(s,t)&=&\frac{2}{ m_{\pi}^2} \lambda_2(s,t) \ ,
   \label{B-coeff}    
\end{eqnarray}
with the generalised polarizability functions as
\begin{eqnarray}
	 \lambda_1(s,t) &=&   - \frac{m_{\pi}^4}{2} 
       \bigg\{ g_{\rho\rightarrow\pi\gamma}^2 \left( 
        \frac{1}{s-m_{\rho}^2}+\frac{1}{u-m_{\rho}^2} \right)   \nonumber \\
  && +  g_{a_1\rightarrow\pi\gamma}^2 \left(\frac{1}{s-m_{a_1}^2}
      +\frac{1}{u-m_{a_1}^2}\right) \bigg\}  \ ,\label{lambda-one}  \\
   \lambda_2(s,t) &=& m_{\pi}^2 \bigg\{
  g_{\sigma\rightarrow \pi\pi} g_{\sigma\rightarrow \gamma\gamma}
   \frac{1}{t-m_{\sigma}^2} 
    -\frac{1}{4} g_{\rho\rightarrow\pi\gamma}^2 \left(\frac{s+m_{\pi}^2}{s-m_{\rho}^2}
    +\frac{u+m_{\pi}^2}{u-m_{\rho}^2}\right) \nonumber\\
   & & + \frac{1}{4} g_{a_1\rightarrow\pi\gamma}^2 
    \left(  \frac{s-m_{\pi}^2}{s-m_{a_1}^2}
      +\frac{u-m_{\pi}^2}{u-m_{a_1}^2} \right) \bigg\} \ . \label{lambda-two}  
\end{eqnarray}

At the pion-Compton threshold  $s=u=m_{\pi}^2$ and $t=0$.  
In chiral-Lagrangian theory the threshold values of the polarizability functions are
$\lambda_1(m_{\pi}^2,0)=0$ and $\lambda_2(m_{\pi}^2,0)=-0.013$.
In the exchange model the threshold functions are dominated by $\sigma$ exchange. 
However, the parameters of the $\sigma$  are rather uncertain and we choose to fix them 
so that  the $\sigma$ contribution to the poarizability functions is twice as large
as the chiral-Lagrangian prediction, and more in agreement with experiment 
\cite{5,5a}. This is further discussed in the Appendix. Thus, the $\sigma$-, $\rho$- and 
$a_1$-exchange contributions  to our threshold polarizability functions are
\begin{eqnarray}
	 \lambda_1(s,t) &=& 0+0.0003+0.0003  \ ,\label{lambda-num-one}  \\
   \lambda_2(s,t) &=& -0.0261+0.0004+0 \ . \label{lambda-num-two}  
\end{eqnarray}
\newpage
\section{Nuclear cross sections}

We are interested in a kinematic region where the transverse momentum components 
of particles can be neglected compared with their longitudinal momentum components. 
Details of the kinematics are given in \cite{PrimakI}.
The cross-section distribution in the pion-nucleus lab system is
\begin{equation}
	\rd \sigma =\frac{1}{4p_1 M_A} \left| {\mathcal M}\right|^2 {\rm dLips} \ ,
	 \label{Cross-nucl-def}
\end{equation}
where $p_1$ is the incident pion lab momentum. The Lorentz-invariant phase space
can be parametrised as
\begin{equation}
{\rm dLips} = \frac{1}{16\pi M_A} \frac{\rd^2 p_{2\bot}}{(2\pi)^2}
    \frac{\rd^2q_{2\bot}}{(2\pi)^2}\frac{\rd q_{2z}}{p_{2z} q_{2z}} \ .
 \label{Phase-space}
\end{equation}

The nuclear Born approximation is represented by the 
one-photon exchange graph is pictured in fig.~3. The small blob in the
\begin{figure}[h]\begin{center}
\scalebox{1.20}{\includegraphics{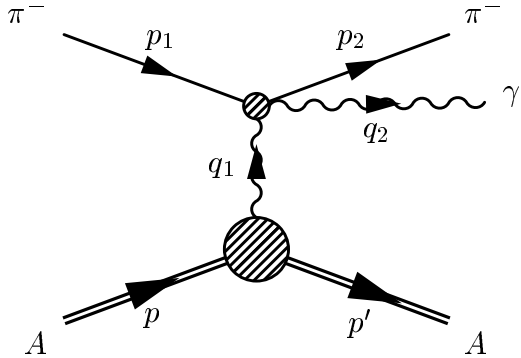}}
\end{center}
\caption{Born diagram for pionic bremsstrahlung.}
\end{figure}
graph represents the full pion-Compton amplitude;
the large blob the photon-nucleus electromagnetic vertex.
The pion charge is $-e$, the nuclear charge $Ze$, and 
the nucleus is treated as a spin-zero particle. 
With $q_1$ the virtual photon four-momentum, these
assumptions lead to a
Coulomb production amplitude  
\begin{equation}
  {\mathcal M}_{B}=\frac{-i}{q_1^2} 
   {\mathcal M}_{\mu\nu}(p_2,q_2;p_1,q_1)(-iZe)
             (p+p')^{\mu}\epsilon_2^{\nu} \ .
\label{Coulomb_Born}
\end{equation}
Since the Compton tensor $ {\mathcal M}_{\mu\nu}$ is 
 gauge invariant we may also make the
replacement $p+p'=2p+q_1\rightarrow 2p$.

The reduction of ${\mathcal M}_{B}$ is much simplified if we
first introduce the parameter
\begin{equation}
x=
  \frac{q_{2z}}{p_1} =\frac{\omega_2}{E_1} \ .  \label{x-fraction}
\end{equation}
Inserting in eq.(\ref{Coulomb_Born}) the expansion of the
pion-Compton amplitude from eq.(\ref{Compt-decomp}) and 
exploiting the techniques of \cite{PrimakI} we get
\begin{eqnarray}
{\cal M}_B&=&\frac{-8\pi iZ M_A e\alpha}{q_1^2} \bigg[ 4E_2 A(s,t) \bigg\{
	 \frac{\mathbf{q}_{2\bot}}{\mathbf{q}_{2\bot}^2 + x^2 m_{\pi}^2 }
	   - \frac{\mathbf{q}_{2\bot}-x\mathbf{q}_{1\bot}} 
	   { (\mathbf{q}_{2\bot}-x \mathbf{q}_{1\bot})^2+ x^2 m_{\pi}^2 }\bigg\}
	      \nonumber \\   &&\nonumber \\
	    &&\quad \quad\qquad\qquad \qquad+\omega_2 B(s,t)\mathbf{q}_{1\bot}
	      \,\bigg]\cdot\bmath{\epsilon}_2 \ .
	 \label{Simple-amp} 
\end{eqnarray}
The subscript $\bot$ indicates a vector component in the impact plane,
{\it i.e.}~the plane orthogonal to the incident momentum $\mathbf{p}_1$, 
which is along the 
$z$-direction. Note that the polarisation vector $\bmath{\epsilon}_2$ is
orthogonal to  $\mathbf{q}_2$, and therefore has both  transverse and 
longitudinal components, the dominant one being transverse. Since
 $\mathbf{p}_{1\bot}=0$ the transverse vector components are related by
\begin{equation}
\mathbf{q}_{1\bot}=\mathbf{p}_{2\bot}+\mathbf{q}_{2\bot} \ .
\label{qorth-sum}
\end{equation}

The second term on the right hand side of eq.(\ref{Simple-amp}) 
has been slightly rewritten as compared with the corresponding term 
in eq.(6.49) 
of \cite{PrimakI}. As a consquence we see directly that the  matrix
element is proportional to  $\mathbf{q}_{1\bot}$, as it should be.

We are interested in hard photons. Therefore, the parameter $x$ of
eq.(\ref{x-fraction}) 
is sizeable, but still in the region $0<x<1$. We are also 
limiting ourselves 
to the Coulomb region, where $q_{1\bot}$ is of the same
size as $q_{1z}$, which is equal to
\[
q_{min}
	  =\frac{m_{\pi}^2\omega_2}{2E_1E_2} = xm_{\pi}(m_{\pi}/2E_2) \ ,
\]
so that $q_{1\bot}\ll m_{\pi}$. The momentum components 
$\mathbf{p}_{2\bot}$
and $\mathbf{q}_{2\bot}$, on the other hand,
 may both be in the GeV region but only in such a way that
their sum $\mathbf{q}_{1\bot}$ remains the size of $q_{min}$. 
It follows that
\begin{equation}
\left| x\mathbf{q}_{2\bot}\cdot \mathbf{q}_{1\bot} \right|
 \ll ( \mathbf{q}_{2\bot}^2 + x^2 m_{\pi}^2 ) \ . \label{inequality}
\end{equation}
Application of this inequality  simplifies the  Born amplitude into 
 \begin{eqnarray}
{\cal M}_B&=&\frac{-8\pi iZ M_A e\alpha}{q_1^2} 
   \frac{4xE_2 }{\mathbf{q}_{2\bot}^2 + x^2 m_{\pi}^2 }
 \bigg[  \tilde{A}(x,\mathbf{q}_{2\bot}^2) 
 \bigg\{\mathbf{q}_{1\bot} - 2\mathbf{q}_{2\bot}
     \frac{\mathbf{q}_{2\bot}\cdot \mathbf{q}_{1\bot}}
          {\mathbf{q}_{2\bot}^2 + x^2 m_{\pi}^2 }  \bigg\}
	      \nonumber \\   &&\nonumber \\
	    &&\quad \quad\qquad\qquad \qquad+ 
	      \tilde{B}(x,\mathbf{q}_{2\bot}^2) \mathbf{q}_{1\bot}
	      \,\bigg]\cdot\bmath{\epsilon}_2 \ ,
	 \label{SimpleII-amp} 
\end{eqnarray}
with 
\begin{eqnarray}
	\tilde{A}(x, \mathbf{q}_{2\bot}^2)&=& A(s,t)  \ , \label{def-tildefcnA}\\
  \tilde{B}(x, \mathbf{q}_{2\bot}^2)&=& 
  \frac{1}{4(1-x)}( \mathbf{q}_{2\bot}^2+x^2m_{\pi}^2) B(s,t) \ .
  \label{def-tildefcnB}
\end{eqnarray}

Here, we have replaced the variables $s$ and $t$ by 
the variables $x$ and $\mathbf{q}_{2\bot}^2$. That this is possible 
follows from a study of the kinematic variables $s$, $t$, and $u$ 
of the virtual pion-Compton scattering, defined in eq.(\ref{Mandelstam}).
 Evaluating them  with the on-shell 
four-momenta $p_1$, $p_2$, and $q_2$, and making use of the 
inequality \ref{inequality} leads to the simple expressions
\begin{equation}
\begin{array}{rcl}
  s - m_{\pi}^2 &=& \displaystyle{ \frac{1}{x(1-x)} }
    \bigg[\mathbf{q}_{2\bot}^2 + x^2 m_{\pi}^2 \bigg] \ ,\\ \\
  t &=&  \displaystyle{\frac{-1}{1-x} }\bigg[\mathbf{q}_{2\bot}^2 + x^2 m_{\pi}^2 \bigg] \ ,\\ \\
  u - m_{\pi}^2&=&  \displaystyle{\frac{-1}{x} }\bigg[\mathbf{q}_{2\bot}^2 + x^2 m_{\pi}^2 \bigg] \ .
\end{array}    
\label{MandelII}
\end{equation}  
We stress that these expressions are valid only for hard bremsstrahlung in the
Coulomb region.
Furthermore, we may on the right hand sides replace 
${q}_{2\bot}^2$ by  ${p}_{2\bot}^2$  without any numerical consequences .

Up to now we have been concerned with the Born approximation, the one-photon
exchange. Including elastic scattering to all orders
induces some changes. The external radiation contributions, corresponding 
to the unit term in $A(s,t)$, comes multiplied by an off-shell 
elastic Coulomb amplitude. In \cite{PrimakI} we were not able to show 
whether the off-shellness gives rise to a phase factor different from the elastic
one. In absence of a firm prediction we assume the phase to be the same as
the elastic one. The polarizability contributions were shown to have 
a form factor being the sum of the elastic Coulomb amplitude plus
an extra term $f_P$. This second contribution is however finite at the
Coulomb peak and does not exhibit the characteristic cross-section peak. It can
be neglected at the our level of accuracy. The same remark applies 
to the nuclear contributions. Adopting these caveats we put
\begin{equation}
	\left|{\cal M}\right|^2  =\left|  {\cal M}_B^2\right| \ .
\end{equation}

The summation over the photon polarisation is trivial. It replaces 
scalar products like $ \left|\mathbf{q}_{2\bot}\cdot\mathbf{\epsilon}_2 \right|^2$
by $ \left|\mathbf{q}_{2\bot} \right|^2$. 
In view of the relation (\ref{qorth-sum}) we may also replace the 
phase-space volume $\rd^2q_{2\bot}  \rd^2p_{2\bot}$ by
$\rd^2q_{1\bot}  \rd^2q_{2\bot}$. 
The cross-section distribution in the pion-nucleus lab system, as defined in 
eq.(\ref{Cross-nucl-def}), then takes the form
\begin{eqnarray}
\frac{\rd \sigma}{\rd^2q_{1\bot}  \rd^2q_{2\bot} \rd x}
  &= &\frac{4Z^2\alpha^3}{\pi^2m_{\pi}^4}
 \Bigg[ \frac{\bmath{q}_{1\bot}^2}{(\bmath{q}_{1\bot}^2+q_{min}^2)^2}\Bigg]
  \Bigg[ \frac{1-x}{x^3} \Bigg] 
   \Bigg[ \left( \frac{x^2 m_{\pi}^2}{\mathbf{q}_{2\bot}^2+x^2 m_{\pi}^2}\right)^2\Bigg] 
         \cdot   \nonumber \\
   &&\Bigg[ \tilde{A}(x,\mathbf{q}_{2\bot}^2)
    \left( \hat{ \mathbf{q}}_{1\bot} - 2\mathbf{q}_{2\bot}
     \frac{\mathbf{q}_{2\bot}\cdot \hat{ \mathbf{q}}_{1\bot}}
          {\mathbf{q}_{2\bot}^2 + x^2 m_{\pi}^2 }  \right)
   \nonumber \\ \nonumber \\
   && \qquad + \tilde{B}(x,\mathbf{q}_{2\bot}^2) \hat{ \mathbf{q}}_{1\bot} \Bigg]^2 \ .
 \label{Coul-peak-cross-general}
 \end{eqnarray}
The scalar functions $\tilde{A}(x, \mathbf{q}_{2\bot}^2)$ and
$\tilde{B}(x, \mathbf{q}_{2\bot}^2)$ are defined in eqs (\ref{def-tildefcnA})
and (\ref{def-tildefcnB}). 
Introducing the functions  $\lambda_1(x, \mathbf{q}_{2\bot}^2)$
and $\lambda_1(x, \mathbf{q}_{2\bot}^2)$ of eqs(\ref{lambda-one}) 
and (\ref{lambda-two}) that describe the non-Born contributions we get
 \begin{eqnarray}
	\tilde{A}(x, \mathbf{q}_{2\bot}^2)&=&  1 -\frac{x^2}{4(1-x)}
	\left( \frac{\mathbf{q}_{2\bot}^2+x^2 m_{\pi}^2}{x^2 m_{\pi}^2}\right)^2
	   \lambda_1(x, \mathbf{q}_{2\bot}^2) \ , \label{Atild-fin} \\
  \tilde{B}(x, \mathbf{q}_{2\bot}^2)&=& 
   \frac{x^2}{2(1-x)}\left(\frac{\mathbf{q}_{2\bot}^2+x^2 m_{\pi}^2}{x^2 m_{\pi}^2}\right)
     \lambda_2(x,\mathbf{q}_{2\bot}^2) \ .\label{Btild-fin} 
\end{eqnarray}

The approximations leading to the cross-section distributions described by 
eqs.(\ref{Coul-peak-cross-general}-\ref{Btild-fin}) demand that the pionic
radiation is hard, that the transverse components of the vectors $\mathbf{q}_2$
and  $\mathbf{p}_2$ are much smaller than their longitudinal components,
and that we are in the Coulomb dominated region where the length 
of the vector $\mathbf{q}_{1\bot}=\mathbf{p}_{2\bot}+\mathbf{q}_{2\bot}$
is of a size similar to that of $q_{1z}=q_{min}$.
 
It is important to realise that although the cross-section distribution in general
depends on the angle
\begin{equation}
\mu = {\hat{ \mathbf{q}}_{1\bot}} \cdot {\hat{ \mathbf{q}}_{2\bot}}=\cos\varphi_{12}\ ,
\label{def-mu}
\end{equation}
the arguments of the polarizability functions $\lambda_1$ and $\lambda_2$ do not.

We end by re-emphasising that $s$, $t$, and $u$ are variables of the virtual 
pion-Compton scattering. The momentum transfer squared to the nucleus is
\begin{equation}
	-t_A=\mathbf{q}_{1\bot}^2+q_{min}^2 \ .
\end{equation}
%
%
\newpage
\section{Under the Coulomb peak: I}

Results concerning the pion polarizabilities are most easily discussed 
if we limit ourselves to the phase-space region where both 
momenta ${p}_{2\bot}$ and ${q}_{2\bot}$ are in the Coulomb region, 
{\it i.e.}~the size of their momenta are of the order of
$q_{min}$ and hence negligible in comparison with $x m_{\pi}$.
In this kinematical situation the variables
$s$, $t$, and $u$ of the pion-Compton amplitude, eq.(\ref{MandelII}),
become simple function of $x$
 \begin{equation}
\begin{array}{rcl}
  s &=&   m_{\pi}^2/(1-x) \ , \\ \\
  t &=&-x^2 m_{\pi}^2/(1-x) \ , \\ \\
  u &=& (1-x) m_{\pi}^2 \ .
\end{array}    
\label{Mandelx}
\end{equation}  

The cross section can be written as
\begin{eqnarray}
\frac{\rd \sigma}{\rd^2q_{1\bot}  \rd^2q_{2\bot} \rd x}
&=& \frac{4Z^2\alpha^3}{\pi^2m_{\pi}^4}
 \Bigg[ \frac{\bmath{q}_{1\bot}^2}{(\bmath{q}_{1\bot}^2+q_{min}^2)^2}\Bigg]
  \Bigg[ \frac{1-x}{x^3} \Bigg] \cdot \nonumber\\
  && \Bigg[ 1+\frac{x^2}{1-x}\left(-\frac{1}{4}\lambda_1(x) + \frac{1}{2} \lambda_2(x)\right)\Bigg]^2
 \ , \label{Coul-peak-cross}
 \end{eqnarray}
 with $x=\omega_2/E_1$ from eq.(\ref{x-fraction}). We conclude that
 in general the contributions from the pion structure terms
 are small. Only if $x$ is very near unity do we get a substantial
 contribution. This means bremsstrahlung photons of energies very
 near those of the incident pion. We also observe that when 
 $x\approx 1$ the energy in the pion-Compton scattering 
 may become so large that the threshold approximation to the 
 polarizability functions breaks down.
 
\begin{figure}[t]\begin{center}
\scalebox{0.40}{\includegraphics{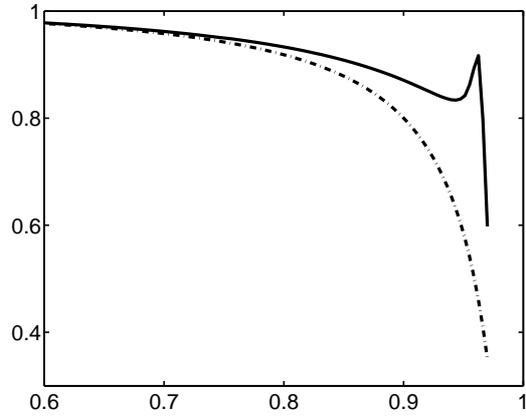}}
\end{center}
\caption{Proportionality function $R(x,\lambda(x))$ in the double Coulomb region. 
   The solid line obtains in the full calculation, the dashed line 
   in the threshold approximation.}\label{COMPASSI}
\end{figure}

 We illustrate the sensitivity to the polarizability functions 
 and their energy dependence by plotting in fig.~\ref{COMPASSI} 
 the proportionality function
 \begin{equation}
  R(x, \lambda(x))=
   \left|1+\frac{ x^2}{1-x} 
   \left(-\frac{1}{4}\lambda_1(x) + \frac{1}{2} \lambda_2(x)\right) \right|^2 \ ,
  \label{R-fcn}
\end{equation}
together with $R(x, \lambda(0))$. The curves are plotted for 
$x$ in the interval (0.6, 0.97) and  
the input parameters are those of the exchange model of Sect.~2. 
It is feasable to extract information about the polarizability functions
only if $R(x, \lambda(x))$ deviates appreciably from unity, which occurs when
 $x$ approaches one. When this happens the curves for $R(x, \lambda(x))$ 
 and $R(x, \lambda(0))$ start to diverge from each other. 
Thus, when the 
 polarizability contributions become appreciable the 
 threshold approximation deteriorates. The structure in the solid curve at 
 $x\approx 0.96$, corresponding to $\sqrt{s}\approx m_\rho$, 
  is caused by the $\rho$-exchange contribution.
\vspace{5cm}
\newpage
\section{Under the Coulomb peak: II}

Our investigation presumes two conditions;  
the transverse components
of the vectors $\mathbf{p}_2$ and $\mathbf{q}_2$ must be much
smaller than their longitudinal components, and
their sum $\mathbf{p}_{2\bot}+\mathbf{q}_{2\bot}=\mathbf{q}_{1\bot}$
must be in the Coulomb region, {\it i.e.}~in the region where
the length of $\mathbf{q}_{1\bot}$ is of a size similar to 
$q_{min}$. If these conditions are not met, we are outside 
the Coulomb peak and nuclear contributions play a role.  
Even though the sum of the two vector components 
$\mathbf{p}_{2\bot}$ and $\mathbf{q}_{2\bot}$ must be very
small, this need not be so for the two vector components individually. 
We shall now consider the case where the two components are
large, which we shall take to mean large in comparison with $m_{\pi}$.

Returning to the general expression (\ref{Coul-peak-cross-general})
we have
\begin{eqnarray}
\frac{\rd \sigma}{\rd^2q_{1\bot}  \rd^2q_{2\bot} \rd x}
  &= &\frac{4Z^2\alpha^3}{\pi^2m_{\pi}^4}
 \Bigg[ \frac{\bmath{q}_{1\bot}^2}{(\bmath{q}_{1\bot}^2+q_{min}^2)^2}\Bigg]
  \Bigg[ \frac{1-x}{x^3} \Bigg] 
   \Bigg[ \left( \frac{x^2 m_{\pi}^2}{\mathbf{q}_{2\bot}^2+x^2 m_{\pi}^2}\right)^2\Bigg] 
         \cdot   \nonumber \\      
   &&\Bigg[ \tilde{A}^2(x,\mathbf{q}_{2\bot}^2)
    \left\{ 1 - \mu^2(1- \kappa^2) \right\} \nonumber \\  
    && \qquad+2\tilde{A}(x,\mathbf{q}_{2\bot}^2)\tilde{B}(x,\mathbf{q}_{2\bot}^2)
    \left\{ 1 - \mu^2(1- \kappa) \right\}\nonumber \\  
    && \qquad+ \tilde{B}^2(x,\mathbf{q}_{2\bot}^2)\Bigg] \ ,
 \label{Coul-peak-crossII}
 \end{eqnarray}
with $\kappa$ a parameter defined as
\begin{equation}
	\kappa(x, \mathbf{q}_{2\bot}^2)=
	  \frac{x^2 m_{\pi}^2- \mathbf{q}_{2\bot}^2}{x^2 m_{\pi}^2 +\mathbf{q}_{2\bot}^2}
	 \ . \label{kappa}
\end{equation}
The angular dependence comes in through
$\mu = {\hat{\mathbf{q}}_{1\bot}} \cdot {\hat{\mathbf{q}}_{2\bot}}$
of eq.(\ref{def-mu}). It is quite simple, and integrating over angles 
means replacing $\mu^2$ by its average, which is $\half$.

In Sect.~4 we considered the region $\mathbf{q}_{2\bot}^2 \ll x^2 m_{\pi}^2$
where the parameter $\kappa=1$ so that the angular dependent terms in 
eq.(\ref{Coul-peak-crossII}) dropped out. Also, the dependence on 
$\mathbf{q}_{2\bot}^2$ in the right hand side of this equation goes away,
leaving a dependence on $x$, and the characteristic point-like Coulomb peak factor 
depending on $\mathbf{q}_{1\bot}^2$.

Now, we consider the region of large transverse momenta where
\begin{equation}
	\mathbf{q}_{2\bot}^2 \gg x^2 m_{\pi}^2 \ ,
	  \label{L-ineq}
\end{equation}
an inequality  equally valid for $\mathbf{p}_{2\bot}^2$. 
The expressions for the kinematic variables in the Compton 
process now simplify, replacing eq.(\ref{MandelII}) by
\begin{equation}
\begin{array}{rcl}
  s &=&  m_{\pi}^2 + \displaystyle{ \frac{\mathbf{q}_{2\bot}^2}{x(1-x)} } \ , \\
  t &=&-\displaystyle{ \frac{\mathbf{q}_{2\bot}^2}{1-x}  }\ , \\
  u &=& m_{\pi}^2 -\displaystyle{ \frac{\mathbf{q}_{2\bot}^2}{x} }\ .
\end{array}    
\label{MandelIII}
\end{equation}  

In the region defined by the inequality (\ref{L-ineq}) we infer from
the definition (\ref{kappa}) that  $\kappa=-1$. As a consequence there 
is no angular dependence in the term proportional to $\tilde{A}^2$.
Furthermore, if we average over angles the cross term proportional 
to  $\tilde{A}\tilde{B}$ vanishes. The term proportional to $\tilde{B}^2$
carries no angular dependence.
Thus, after integration over 
angles the cross-section distribution becomes
 \begin{eqnarray}
\frac{\rd \sigma}{\rd q_{1\bot}^2  \rd q_{2\bot}^2 \rd x}
  &= &\frac{4Z^2\alpha^3}{\bmath{q}_{2\bot}^4} 
 \Bigg[ \frac{\bmath{q}_{1\bot}^2}{(\bmath{q}_{1\bot}^2+q_{min}^2)^2}\Bigg]
  x(1-x)\ \nonumber\\
 && \cdot\Bigg[ \tilde{A}^2(x, \mathbf{q}_{2\bot}^2) 
   + \tilde{B}^2(x, \mathbf{q}_{2\bot}^2) \Bigg] \ .
 \label{Cross-large-q:s}
 \end{eqnarray}
The invariant functions entering this equation are defined in 
eqs (\ref{def-tildefcnA},\ref{def-tildefcnA}) and eqs (\ref{A-fac0},\ref{B-fac0})
and become
\begin{eqnarray}
	\tilde{A}(x, \mathbf{q}_{2\bot}^2) &=& 1 -\frac{1}{4x^2(1-x)}
	  \left(\frac{\mathbf{q}_{2\bot}^2}{m_\pi^2}\right)^2 \lambda_1(x, \mathbf{q}_{2\bot}^2) \ ,\\
	\tilde{B}(x, \mathbf{q}_{2\bot}^2) &=& \frac{1}{2(1-x)}
	\left(\frac{\mathbf{q}_{2\bot}^2}{m_\pi^2}\right) \lambda_2(x,\mathbf{q}_{2\bot}^2) \ .
\end{eqnarray}
We notice that the polarizability contributions are enhanced by powers of 
the factor $\mathbf{q}_{2\bot}^2/m_\pi^2$.

Finally, in the phase-space element of the cross-section distribution
(\ref{Cross-large-q:s}) we may introduce the momentum transfer 
to the nucleus $t_A$ via the  identity
\begin{equation}
\rd q_{1\bot}^2  = -\rd t_A  \ .
\end{equation}

We illustrate the sensitivity of the cross-section distributions to the polarizability
functions  and their energy dependences in the same way as we did 
for small transverse momenta in Sect.~4.
Thus,  introduce the proportionality function  
 \begin{equation}
  R(x, \mathbf{q}_{2\bot}^2; \lambda(x,\mathbf{q}_{2\bot}^2))=
   \Bigg[ \tilde{A}^2(x, \mathbf{q}_{2\bot}^2) 
   + \tilde{B}^2(x, \mathbf{q}_{2\bot}^2) \Bigg] \ ,
    \label{R2-fcn}
\end{equation}
which now depends on both $x$ and $\mathbf{q}_{2\bot}^2$. 
Putting  $\lambda_1$ and $\lambda_2$ equal to zero leads to $R=1$, 
the value for a point-like pion. 

\begin{figure}[ht]\begin{center}
\scalebox{0.40}{\includegraphics{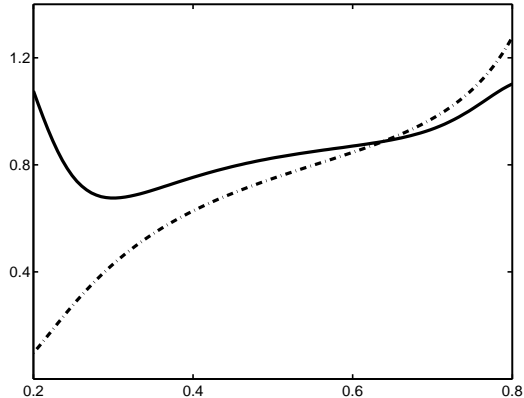}}
\end{center}
\caption{Proportionality function 
  $R(x,\mathbf{q}_{2\bot}^2; \lambda(x,\mathbf{q}_{2\bot}^2))$ 
  for $q_{2\bot}=3.5 m_\pi$ in the single Coulomb region. 
   The solid line obtains in the full calculation, the dashed line 
in the threshold approximation.}\label{COMPASSII}
\end{figure}

In fig.~\ref{COMPASSII} we graph the proportionality function
$R(x, \mathbf{q}_{2\bot}^2; \lambda(x,\mathbf{q}_{2\bot}^2))$ and
the  function  $R(x, \mathbf{q}_{2\bot}^2; \lambda(0))$, which is
the same function but evaluated  with the threshold values of the 
polarizabilities. 
For the illustration we have chosen $q_{2\bot}=3.5 m_\pi$. 
From eq.(\ref{MandelIII}) it then follows  that $s\geq 50m_\pi^2$
and far from its threshold value $s=m_\pi^2$. The curves are plotted for $x$ in the 
region $0.2\leq x\leq 0.8$, corresponding to $\sqrt{s}$ in the region
$990\leq \sqrt{s}\leq 1230$, in units of MeV. In view of the large energies
it is not astonishing to  realise that  the threshold
approximation is unrealistic. The solid curve which represents the full calculation
is not symmetric around $x=0.5$ even though the energy $s$ is. The reason is that 
neither $t$ nor $u$ is symmetric. Therefore, the cross-section distributions at $x=0.4$ 
and $x=0.6$, {\itshape  e.~g.},  measures the pion-Compton cross-section distribution
at completely different scattering angles. But it is futile attempting to extract values of
the functions $\tilde{A}(x, \mathbf{q}_{2\bot})$ and $\tilde{B}(x, \mathbf{q}_{2\bot})$ 
which are both complex.
%
%
%
%
\newpage
\section{Summary}

Hard  bremsstrahlung in high-energy pion-nucleus scattering  
 in the Coloumb region has been investigated. 
 The kinematics of the reaction can be read off from
\[
\pi^-(p_1)+A(p)\rightarrow \gamma(q_2)+ \pi^-(p_2) +A(p') \ .
\]
The resctriction to the Coulomb region means that
 the momentum transfer to the nucleus is of the order of
$q_{min}
	  =m_{\pi}^2\omega_2/(2E_1E_2)$. 
As a consequence, the production amplitude is dominated by the one-photon
exchange diagram and the cross-section distribution exhibits the 
well-known  Coloumb-production peak structure. 
In the Coloumb region the sum of the tranverse 
momenta  
\[
\mathbf{p}_{2\bot}+\mathbf{q}_{2\bot}=\mathbf{q}_{1\bot}  \ ,
\]
is tiny, although the transverse momenta themselves need not be that small.

We have derived an expression for the cross-section distribution,
eq.(\ref{Coul-peak-crossII}),
valid when the transverse momenta of the emerging photon and pion
are much smaller than their longitudinal momenta.
The arguments of this expression are $x= \omega_2/E_1$ and $\mathbf{q}_{2\bot}^2$, which for
all  practical purposes is the same as $\mathbf{p}_{2\bot}^2$. The nuclear
production amplitude involves the on-shell pion-Compton amplitude at energies and 
angles that depend on the values of $x$ and $\mathbf{q}_{2\bot}^2$, eq.(\ref{MandelII}).
The pion-Compton amplitude is modelled as a sum of the point-like Born terms 
and the polarizability terms, represented by $\sigma$, $\rho$, and $a_1$ 
exchange-diagram terms.

We have illustrated our model by considering two limits; in the first one,
the double Coloumb region, the
sizes of both $|\mathbf{q}_{2\bot}|$ and $|\mathbf{p}_{2\bot}|$ are of the order
$q_{min}$, whereas in the second one, the single Coloumb region, 
their sizes are both much larger than $m_\pi$,
but in such a way that their vector sum remains small.

In the double-Coloumb region the pion-Compton amplitude depends only on $x$. 
When $x$ is small the influence of the polarizability functions is 
weak. In order to be noticed we must go to $x$-values near
unity. Then, the energy in the pion-Compton becomes so large that the
threshold approximation of the polarizability functions becomes questionable.
To extract reliable values for the famous coefficients 
$\alpha_\pi$ and  $\beta_\pi$ requires accurate experiments.

In the single-Coloumb region the effective pion-Compton energies are several times larger 
than the pion mass and the threshold approximation to the polarizability functions
becomes unrealistic. Moreover, the polarizability functions develop imaginary parts that are
important. Furthermore, the nuclear cross-section distribution is not related
in a simple way to a pion-Compton distribution at a fixed energy, 
since the energy $s$ and momentum transfer $t$ in the pion-Compton scattering 
are functions of $x$ and $\mathbf{q}_{2\bot}^2$.

%

%
%
%
%
%
\newpage
\section{Appendix}
In this Appendix the parameters of the pion-Compton model are discussed.

The radiative decay of the sigma meson has width and coupling constant
related by
\begin{equation}
	\Gamma(\sigma\rightarrow\gamma\gamma)=\pi\alpha^2 
	 g_{\sigma\rightarrow\gamma\gamma}^2 m_{\sigma}^3 \ .
\end{equation}
For the strong decay of the sigma meson the corresponding relation is
\begin{equation}
	\Gamma(\sigma\rightarrow\pi^+\pi^-)=\frac{1}{16\pi m_{\sigma}}
 g_{\sigma\rightarrow\pi\pi}^2 \sqrt{ 1-\frac{ 4m_{\pi}^2}{m_{\sigma}^2}}\ .
\end{equation}
Numerical values for the coupling constants have been extracted
by Fil'kov and Kashevarov \cite{Fil} from a study of data for the reaction
$\gamma\gamma\rightarrow \pi^0\pi^0$. 
Their results are: $ \Gamma(\sigma\rightarrow\gamma\gamma)=0.62$ keV, 
$\Gamma(\sigma\rightarrow\pi^+\pi^-)=803$ MeV, and $m_{\sigma}=547$ MeV.
For the product of coupling constants these numbers give
$g_{\sigma\rightarrow\gamma\gamma}g_{\sigma\rightarrow\pi\pi}  = 0.762$,
which results in a value for $\beta_{\pi^+}$ almost four times as large as 
the chiral-Lagrangian prediction. In view of the uncertainty of the
$\sigma$ parameters we shall choose
\begin{equation}
		g_{\sigma\rightarrow\gamma\gamma}
	g_{\sigma\rightarrow\pi\pi}  = 0.400 \ .
\end{equation}
giving a value more in line with experimental prejudices \cite{5,5a}. 

The $s$-channel propagators of the $\rho$ and $a_1$ mesons are given 
Breit-Wigner shapes
\begin{eqnarray}
	B^L(s)&=&\frac{m_0^2}{(m_0^2-s) -im_0 \Gamma^L(s)} \ , \\
	\Gamma^L(s)&=& \Gamma_0\left(\frac{k}{k_0}\right)^{2L+1}
	    \frac{m_0}{\sqrt{s}}  \,                                                                             \theta(\sqrt{s}-m_1-m_2) \ ,
\end{eqnarray}
where $\Gamma_0$ and $m_0$ are nominal values and $k_0$ the decay momentum
at mass $m_0$. 
The momentum is
\begin{equation}
	k(s)=\frac{1}{2m_0}\left[(s-(m_1+m_2)^2)(s-(m_1-m_2)^2)\right]^{1/2} \ .
\end{equation}
Furthermore, $L=1$ for $\rho\rightarrow\pi\pi$ whereas $L=0$ for 
 $a_1\rightarrow\rho\pi$. The total nominal widths are 
 $ \Gamma_\rho=150$ MeV, $\Gamma_{a_1}=450$ MeV, and the masses are
$m_{\pi}=139.6$ MeV,  $m_{\rho}=775$ MeV, $m_{a_1}=1230$ MeV.

The relations between width and coupling constant for the $\rho$ meson is
\begin{equation}
\Gamma(\rho^+\rightarrow\pi^+\gamma)= \third \alpha g_{\rho\rightarrow\pi\gamma}^2
   \left[ \frac{m_{\rho}^2-m_{\pi}^2}{2m_{\rho}}\right]^3 \ .
\end{equation}
With  a numerical value for the width $\Gamma(\rho^+\rightarrow\pi^+\gamma)=68$ 
keV the coupling constant becomes
\begin{equation}
	m_\rho g_{\rho\rightarrow\pi\gamma}=0.5644 \ .
\end{equation}

The relations between width and coupling constant for the $a_1$ meson is
\begin{equation}
\Gamma(a_1^+\rightarrow\pi^+\gamma)= \third \alpha g_{a_1\rightarrow\pi\gamma}^2
   \left[ \frac{m_{a_1}^2-m_{\pi}^2}{2m_{a_1}}\right]^3 \ .
\end{equation}
This relation is exactly the same as the one for the $\rho$ meson even though
the parities of the $\rho$ and $a_1$ mesons differ.
With  a numerical value for the width $\Gamma(a_1^+\rightarrow\pi^+\gamma)=640$
 keV the coupling constant becomes
\begin{equation}
	m_{a_1}g_{a_1\rightarrow\pi\gamma}=1.334 \ .
\end{equation}

%
%
%

\end{document}